\documentclass[12pt]{article}
\usepackage{graphicx}
\usepackage{cite}
\def\be{\begin{equation}}
\def\ee{\end{equation}}
\def\ba{\begin{eqnarray}}
\def\ea{\end{eqnarray}}

\def\bq{\begin{quote}}
\def\eq{\end{quote}}

 at 10truept

\newcommand{\beq}{\begin{equation}}
\newcommand{\eeq}{\end{equation}}
\newcommand{\beqa}{\begin{eqnarray}}
\newcommand{\eeqa}{\end{eqnarray}}

\def\lesssim{~\mbox{\raisebox{-.6ex}{$\stackrel{<}{\sim}$}}~}

\def\ltap{\ \raise.3ex\hbox{$<$\kern-.75em\lower1ex\hbox{$\sim$}}\ }
\def\gtap{\ \raise.3ex\hbox{$>$\kern-.75em\lower1ex\hbox{$\sim$}}\ }
\def\gl{\ \raise.5ex\hbox{$>$}\kern-.8em\lower.5ex\hbox{$<$}\ }
\def\roughly#1{\raise.3ex\hbox{$#1$\kern-.75em\lower1ex\hbox{$\sim$}}}

\begin{document}

\thispagestyle{empty}
\begin{flushright}
{\tt astro-ph/0409596}\\
\end{flushright}

\vskip1.5cm
\begin{center}
{\LARGE{\bf Exorcising $w < -1$}}\\
\vskip1.5cm {\large Csaba Cs\'aki$^{a,}$\footnote{\tt
csaki@lepp.cornell.edu}, Nemanja Kaloper$^{b,}$\footnote{\tt
kaloper@physics.ucdavis.edu} and John
Terning$^{b,c,}$\footnote{\tt
terning@lanl.gov}}\\

\vspace{.5cm}

$^a${\em Newman Laboratory of Elementary Particle Physics, Cornell
University}\\
{\em  Ithaca, NY 14853}\\

\vskip 0.1in

$^b${\em Department of Physics, University of California}\\
{\em Davis, CA 95616}\\

\vskip 0.1in

$^c${\em Theoretical Division T-8, Los Alamos National
Laboratory}\\
{\em Los Alamos, NM 87545}\\
\vskip 0.1in \vskip 0.1in
\end{center}
\vskip .25in
We show that the combined dimming of Type Ia supernovae induced by
both a cosmological constant and the conversion of photons into
axions in extra-galactic magnetic fields can impersonate dark
energy with an equation of state $w < -1$. An observer unaware of
the presence of photon-axion conversion would interpret the
additional dimming as cosmic acceleration faster than that induced
by a cosmological constant alone. We find that this mechanism can
mimic equations of state as negative as $w \simeq - 1.5$. Our model does
not have any ghosts, phantoms and the like. It is fully consistent
with the conventional effective field theory in curved space, and all
existing observational constraints on the axions are obeyed.

\vfill \setcounter{page}{0} \setcounter{footnote}{0}
\newpage

\setcounter{equation}{0} \setcounter{footnote}{0}

Cosmological observations suggest that the expansion of the
universe may have begun accelerating, entering a very late stage
of inflation \cite{sne,Riess,cmb,krtur}. If this is true, then the
universe ought be dominated by a dark, non-clumping, energy
component, comprising as much as $70\%$ of the critical energy
density. In order to account for the cosmic acceleration the dark
energy should have negative pressure, obeying roughly $w = p/\rho
< -2/3$ \cite{eqst,maor}. The ``usual suspects" for dark energy
are either a tiny cosmological constant or a time-dependent
quintessence field \cite{q,ams}. However both require ample fine
tunings to fit the data \cite{weinberg}: an unnaturally tiny
energy density, of the order of the current critical energy
density of the Universe, $\rho_c \sim 10^{-12}$ eV$^4$, and in the
case of quintessence a tiny mass $m_Q$ smaller than the current
Hubble parameter $H_0 \sim 10^{-33} eV$ and couplings to the
Standard Model matter weaker than gravity \cite{carroll}. In the
few examples which are consistent with the 4D effective field
theory, quintessence is a pseudo-scalar axion-like field
\cite{axq}, whose symmetries stabilize its mass and render its
couplings to the visible matter sufficiently weak.

At present, the most sensitive probe of the nature of dark energy
are the Type Ia supernovae (SNe) \cite{sne}. Among all the sources
of cosmological data they give us the best bounds on the dark
energy equation of state $w = p/\rho$. It is thus reasonable to
ask if those observations imply that the universe must be
accelerating, or if there might be other explanations. On the
other hand, the improved observations of the CMB and the large
scale structure are now beginning to strengthen the case for the
accelerating universe, even without resorting to the SNe data
\cite{hanmo,max,uros}. Still the strongest bounds on the equation
of state arise only after the SNe observations are included as
well. More curiously, the analysis indicates that the values of $w
< -1$ are allowed \cite{caldwell,ckw,mmot}. This is very puzzling,
since in a four dimensional cosmology based on Einstein's gravity,
this would normally appear to imply violations of the dominant
energy condition, $|p|\le \rho$, long held to be an important
avatar of stability in General Relativity. Indeed, the simplest
models that yield such behavior employ scalar fields with negative
kinetic terms, or ghosts \cite{caldwell,ckw,gibbons,more}. They
have been dubbed phantoms \cite{caldwell}. These models are in
fact just an incarnation of the so-called superexponential
inflation at a much lower scale, discussed originally by Pollock
in 1985 \cite{pollock}. These naive phantom models are plagued
with instabilities. Quantum-mechanically, they are deeply
problematic because they do not admit a stable ground state
\cite{gibbons,cht,wise}. Instabilities also persist at the
classical level \cite{gibbons}, and in general there are
difficulties in trying to make sense of these models within the
framework of effective field theory\footnote{Notice that the ghost
condensation mechanism of \cite{nimarc} would not appear to help
the phantom much. Once the ghosts condense one would end up with
cosmic acceleration driven by a cosmological constant-like term.}
\cite{cht}. While quantum effects may violate the dominant energy
condition, the violations do not appear at phenomenologically
relevant scales \cite{owoo}.

Should we therefore take any indications of $w<-1$ seriously? The
current data do not support this very strongly
\cite{hanmo,max,uros}. There are also claims that different
averaging procedures are needed, possibly removing the support for
$w < - 1$ altogether \cite{wang}. Moreover, it has been argued
\cite{maor} that a variable $w \ge - 1$ is degenerate with a
constant $w < -1$ within the current data sets. The debate on
whether or not $w$ may have dipped below $-1$ at low redshifts
still continue \cite{starobinsky,huterer,bergstrom}. Thus although
the case for $w<-1$ is very weak, an attitude that an observer
should keep an open mind to possibilities and ignore theorist's
prejudice is a healthy one, and search strategies for $w<-1$ have
been proposed \cite{kabri}. But if the data at some future time
really does support $w<-1$, what could this mean? A phantom would
run counter to the established rules of effective field theory,
and so if the cosmological data really support $w<-1$, would it
mean that we have to give up the usual effective field theory
altogether?

One possible way to accept $w < -1$ and circumvent the problems with
instabilities could be to weaken gravity in the far
infrared\footnote{We thank G. Gabadadze and R. Scoccimarro for the
discussions of this issue.}. If the gravitational coupling becomes
weaker at large distances, objects far away would move faster than
they would have been moving if the coupling remained constant. An
observer using the standard Friedmann equation to measure cosmic
distances could interpret this as an effect of cosmic
acceleration, and think of the agent driving acceleration as dark
energy obeying $w<-1$, even though no such agent is really there.
This approach has been pursued in the framework of generalized
scalar-tensor gravity \cite{cdft}, and in the framework of DGP
gravity \cite{dgp} where not only the gravitational constant but
the whole structure of gravity changes at large distances
\cite{lue}. In the former case, arranging for an illusion of
$w<-1$ is hard, and the effects are small. In the latter case, it
appears that a transient stage with $w<-1$ is generic, but the
effects are still not very large\footnote{We thank A. Lue for the
discussion of this issue.}. 
More importantly, in all of these models, and also in the case of
the more naive phantom cosmologies, the effects mimicking $w<-1$
are embedded in gravity and geometry, and so they would affect
everything in the universe {\it equally}, because of the
equivalence principle. Yet, a look at the data shows that at
present it is mainly SNe on their own that might suggest $w<-1$.
All other observations are consistent with $w \ge -1$.

In this article we will pursue a different approach. It is a
variation of our recent proposal \cite{ckt} that the observed
dimming of the SNe arises as a {\it combination} of a faster
expansion of the universe and conversion of photons emitted by SNe
into ultralight axions. Photon-axion conversion will occur in
external magnetic fields, along the lines of the mechanism
exhibited in \cite{rafsto}. We found that it could contribute to
the observed dimmer SNe if the photon-axion coupling scale is
$M\sim 4 \cdot 10^{11}$ GeV and the axion mass is $m\sim 10^{-16}$
eV, assuming intergalactic magnetic fields $B\sim 5 \cdot 10^{-9}$
G coherent over distances of the order of a Mpc, all in agreement
with observational bounds \cite{bound,kronberg}. In our mechanism
the depletion of the photon flux due to the axion production
automatically saturates at about a third of its value at the
source, thanks to the random orientation of the extragalactic
magnetic fields. In this way, we were able to significantly relax
the bounds on the equation of state of the dark energy. While dark
energy still comprised $70\%$ of the contents of the universe, to
explain the observed dimming of SNe together with the photon-axion
conversion we needed the equation of state to be only as negative
as $w = -1/3$. The combined effect of the expansion driven by it
and the photon-axion conversion weakening the flux of photons from
SNe by 1/3 then conspired to impersonate precisely the effects of
the cosmological constant alone \cite{ckt}.

Imagine now that the dark energy is really a cosmological
constant. This is automatically consistent with the CMB and large
scale surveys. However, enter the same axion as the one we
considered in \cite{ckt}. While the CMB and the large scale
structure observations would remain completely unaffected by it,
the optical sources, and in particular SNe would appear dimmer
than they would be in the presence of the cosmological constant
alone! An unsuspecting observer could thus conclude that the
universe must be accelerating even faster than it would have if
the dark energy equation of state obeyed $w\ge -1$, and may be
tempted to resort to phantoms as an explanation. Yet, there is not
a single trace of any supernatural degrees of freedom here - it is
merely the ultralight axion, and the conversion of photons into it
with hardly a trace of it left, which impersonates the
phantom-like effects. In fact, we will show that the combination
of photon-axion conversion and cosmological constant can mimic
phantom-like dark energy with the equation of state as negative as
$w\simeq -1.5$, coming very close to the lower bounds quoted in
\cite{hanmo,max,uros}, and scanning the full interval between $-1$
and $-1.5$. This phenomenon affects bright objects differently
than it does the dark ones: the extra dimming is only manifest
with the sources of electromagnetic radiation and does not affect
dark matter at all. Any attempt to accomplish this with
modifications of the gravitational sector would clearly require
violations of the equivalence principle.  A key aspect of the
mechanism is that since there are no negative contributions to the
Hamiltonian, the theory is perfectly well behaved, and has a
stable vacuum. This explanation is firmly rooted in the realm of
simple effective field theory.

Let us now briefly review the key ingredients of the photon-axion
conversion mechanism \cite{ckt,cktplasma}. The operator governing
photon-axion coupling is
\be
{\cal L}_{int}=
\frac{a}{M} \vec E \cdot \vec B \, ,
\label{coupling}
\ee
where the scale $M$ parameterizes the strength of the axion-photon
interactions. In the extragalactic magnetic fields this induces a
bilinear mixing between the photon and the axion \cite{rafsto}. At
distances small compared to the size of a typical magnetic domain
($\sim$ MPc), a photon whose electric field is orthogonal to $\vec
B$ remains unaffected by this mixing, but a photon whose electric
field is parallel to $\vec B$ mixes with the axion, and these two
flavors oscillate into each other during propagation. Small
variations in the magnetic field absorb the (tiny) rest mass
difference between them. If the line of sight is along the
$y$-direction, and the magnetic field along the $z$-axis, the
field equations are
\be \Bigl\{ \frac{d^2}{dy^2} + {\cal E}^2 - \pmatrix{ \omega_p^2 &
i {\cal E} \frac{B}{M_{L}} \cr -i {\cal E} \frac{B}{M_{L}} & m^2
\cr} \Bigr\} \pmatrix{ |\gamma\rangle  \cr |a\rangle \cr} = 0 \, .
\label{frice} \ee
We have used the Fourier-transforms of the fields in the energy
picture ${\cal E}$ and defined the state-vectors $|\gamma\rangle$
and $|a\rangle$ for the parallel photon and the axion. Here $B =
\langle \vec e \cdot \vec B\rangle \sim | \vec B|$ is the averaged
projection of the extra-galactic magnetic field on the photon
polarization $\vec e$. When the photons propagate in the ionized
interstellar medium, they couple to the sonic wave in the charged
plasma, and acquire an effective mass term $\omega_p^2$, where
$\omega_p$ is the plasma frequency \cite{rafsto}. It is given by
$\omega_p^2=4\pi \alpha n_e/m_e$, with $n_e$ the electron density,
$m_e$ the electron mass, and $\alpha$ the fine structure constant.

Because the mixing matrix in Eq.~(\ref{frice}),
\begin{equation}
{\cal M} = \pmatrix{ \omega_p^2 & i {\cal E} \mu \cr -i {\cal E}
\mu & m^2 \cr} \, , \label{massm}
\end{equation}
where $\mu = B/M$, is  not diagonal, interaction eigenstates $|
\gamma \rangle, | a \rangle$ oscillate into each other. Defining
the propagation eigenstates $|\lambda_-\rangle$ and
$|\lambda_+\rangle$ \cite{ckt}, which diagonalize the matrix
(\ref{massm}), with eigenvalues $\lambda_\mp = \frac{\omega_p^2 +
m^2}{2} \mp \sqrt{\frac{(\omega_p^2 - m^2)^2}{4} + \mu^2 {\cal
E}^2} $, we can solve the Schr\" odinger equation (\ref{frice}).
The survival probability $P_{\gamma \rightarrow \gamma} = |\langle
\gamma(y_0)|\gamma(y)\rangle|^2$ of the photon interaction
eigenstate which travelled a distance $\Delta y$ is then
\cite{ckt,rafsto}
\be P_{\gamma \rightarrow \gamma} = 1 - \frac{4 \mu^2 {\cal
E}^2}{(\omega_p^2 - m^2)^2+4\mu^2 {\cal E}^2} \sin^2\left[
\frac{\sqrt{(\omega_p^2 - m^2)^2 + 4 \mu^2 {\cal E}^2}}{4{\cal E}}
\Delta y\right] \, . \label{proba} \ee

From this formula we see that in the limit ${\cal E} \gg
(m^2+\omega^2_p)/\mu$, the mixing is maximal, and the oscillation
length is practically independent of the photon energy. Hence
high-energy photons will oscillate achromatically. On the other
hand, in the low energy limit ${\cal E} \ll (m^2+\omega^2_p)/\mu$,
the mixing is small, and the oscillations are very dispersive, due
to the energy-dependence of both the mixing angle and the
oscillation length. To decide which of these limits is appropriate
we must turn to the selection of realistic parameters for the
scales in the problem.

We will assume that the averaged value of $\vec B$ is close to its
observed upper limit, and take for the magnetic field amplitude
$|\vec B| \sim 5 \cdot 10^{-9}$ G \cite{kronberg}. To avoid
affecting the small primordial CMB anisotropy, $\Delta T/T \sim
10^{-5}$, we choose the axion mass $m$ to be large enough for the
mixing between microwave photons and the axion to be small. In
order to maximize the couplings of optical range photons, we take
the mass scale $M$ to be as low as possible to remain allowed by
the current bounds on ultralight axions \cite{bound}. These
parameters are in the range of
\be m \sim 10^{-16} {\rm eV} \, , ~~~~~~~~~ M \sim 4\cdot 10^{11}
{\rm GeV} \, . \label{finscale} \ee
These scales maximize the couplings of optical photons and cut off
the mixing in the microwave range. While at early times the CMB
photons were much more energetic, the mixing then was cut off
because there were no sizeable extra-galactic magnetic fields yet,
since their origin is likely tied to structure formation
\cite{egmf}.

So far we have been discussing the evolution of the photon-axion
system inside a coherent magnetic domain. However the cosmological
magnetic fields aren't uniform. Taking $L_{dom} \sim $ Mpc for the
size of a typical coherent magnetic domain \cite{kronberg}, we
solve  for the quantum mechanical evolution of
unpolarized light in a distribution of magnetic domains with
random field directions along the line of sight. Analytic
considerations show that for maximal mixing, when $\cos (\mu
L_{dom})> -1/3$, the photon survival probability is monotonically
decreasing:
\be P_{\gamma \rightarrow \gamma} = \frac{2}{3}+\frac{1}{3} e^{-
\Delta y/L_{\rm dec}} \, . \ee
The decay length is given by \cite{ckt}
\be L_{\rm dec}= \frac{L_{\rm dom}}{ \ln \left( \frac{4}{1+3\cos
(\mu L_{\rm dom})}\right)}~. \ee
For $\mu L_{\rm dom} \ll 1$ this reduces to
\be L_{\rm dec}= \frac{8}{3 \mu^2 L_{\rm dom}}~. \ee
After the voyage through many magnetic domains the initial system
of unpolarized photons undergoes equilibration between the two
photon polarizations and the axion. Hence on average a third of
all photons will become axions after a long trip through many
magnetic domains.

This effect contributes to the total dimming of distant sources
because we should account for the loss of luminosity due to the
axion production. We should replace the absolute luminosity ${\cal
L}$ by an effective one, taking into account the photon survival
rate:
\be {\cal L}_{eff} = {\cal L} ~P_{\gamma \rightarrow \gamma} \, .
\ee
As a result of $P_{\gamma \rightarrow \gamma} < 1$, the effective
luminosity distance determined from ${\cal L}_{eff}$ will appear
larger than the actual distance to the source.

Various aspects of the photon-axion conversion mechanism were
investigated in \cite{jecg,others,mbg}. The most important effects
arise from the presence of the photon plasma mass. This yields a
weak frequency dependence of the dimming, which would be a direct
signature of the mechanism, while still being unobservable on the
current SNe samples \cite{cktplasma,jecg,others,mbg}. It is
reasonable \cite{cktplasma} to imagine that over large fractions
of space at redshifts $z \lesssim 1$ the electron density is at
most $n_e \leq 6 \cdot 10^{-9} {\rm cm}^{-3}$, and possibly even
less than that. This value of the electron density yields the
plasma frequency $\omega_p \leq 3\cdot 10^{-15}$ eV. In this
regime the frequency dependence of the supernova dimming remains
below the current experimental sensitivity. The intergalactic
plasma relaxes the lower bound on $m$ in (\ref{finscale})  since
the plasma-generated effective photon mass by itself has the right
magnitude to suppress the photon-axion mixing at sub eV energies.

More recently, in \cite{mogoo} the authors looked at the frequency
dependence of the spectra of quasi-stellar objects (QSOs) in the
SDSS early data release \cite{sloan}, with the conclusion that QSO
spectra limit the maximal amount of dimming at $z=0.8$ to be less
than about a $0.15$ increase in magnitude. It would be interesting
to repeat this with the QSOs in the new SDSS data release, which
has appeared since. It would also be interesting to see the
effects of allowing the coherence length of the magnetic fields to
be smaller than MPc by only a factor of a few. One then could not
average over the phase factor in the mixing probability (see eq.
(\ref{proba}) below), which might suppress the frequency
dependence. Nevertheless here we will just adopt the constraint
from \cite{mogoo} and show that even with this bound in force
there can still be very important observable effects coming from
photon-axion conversion.

In \cite{baku} the authors have pursued the violations of the
reciprocity relation between the luminosity distance and
angular-diameter distance, using the SNe Ia data to measure the
luminosity distance and the FRIIb radio galaxy data \cite{daly} to
measure the angular-diameter distance. Their bound on photon-axion
mixing can be stated as $(L_{dec} H_0)^{-1}< 0.6$. This is just
slightly stronger (but of similar magnitude) than the bound from
the QSO spectra. However, in a recent paper \cite{uam} different
data was used to test the reciprocity relation. The authors of
\cite{uam} have used the data from X-ray and SZ observations of
clusters, and their analysis finds a weak support for a violation
of the reciprocity relation between the luminosity distance and
angular-diameter distance. This would be consistent with the
photon-axion mixing. In addition it is difficult to quantify the
systematic errors  due to evolution effects  in structures formed
at different redshifts. Therefore in our analysis we will only be
adopting the constraint from \cite{mogoo}. However any
modifications of the results presented here that might ensue from
the constraint of \cite{baku} would not be significant.

Hence it is fair to say that to date the possibility that the
photon-axion conversion contributes to the observed SNe
dimming~\cite{sne} has not been ruled out. In fact, as we will see
later on, apart from the bounds from~\cite{mogoo,baku}, the
original proposal~\cite{ckt} would still provide a very good fit
to the most recent SNe data, including the farthest points and not
conflicting with the WMAP bounds. Here however we will not assume
that the dimming is dominated by photon-axion conversion. Instead
we will imagine that there is a cosmological constant (or some
form of dark energy with $w$ near -1) providing the bulk of the
dimming, while the photon-axion conversion gives an additional
contribution. This way the relative contribution of the
photon-axion conversion to the total dimming of supernovae is
reduced with respect to the case of dark energy with $w=-1/3$
which we discussed previously \cite{ckt}. Therefore the axion
side-effects will be reduced accordingly, further weakening the
already weak frequency dependence. Thus, as we will see, many
interesting examples can be brought in accord with the bounds
from~\cite{mogoo,baku}.
\begin{figure}[thb]
\centerline{\includegraphics[width=0.9\hsize,angle=0]{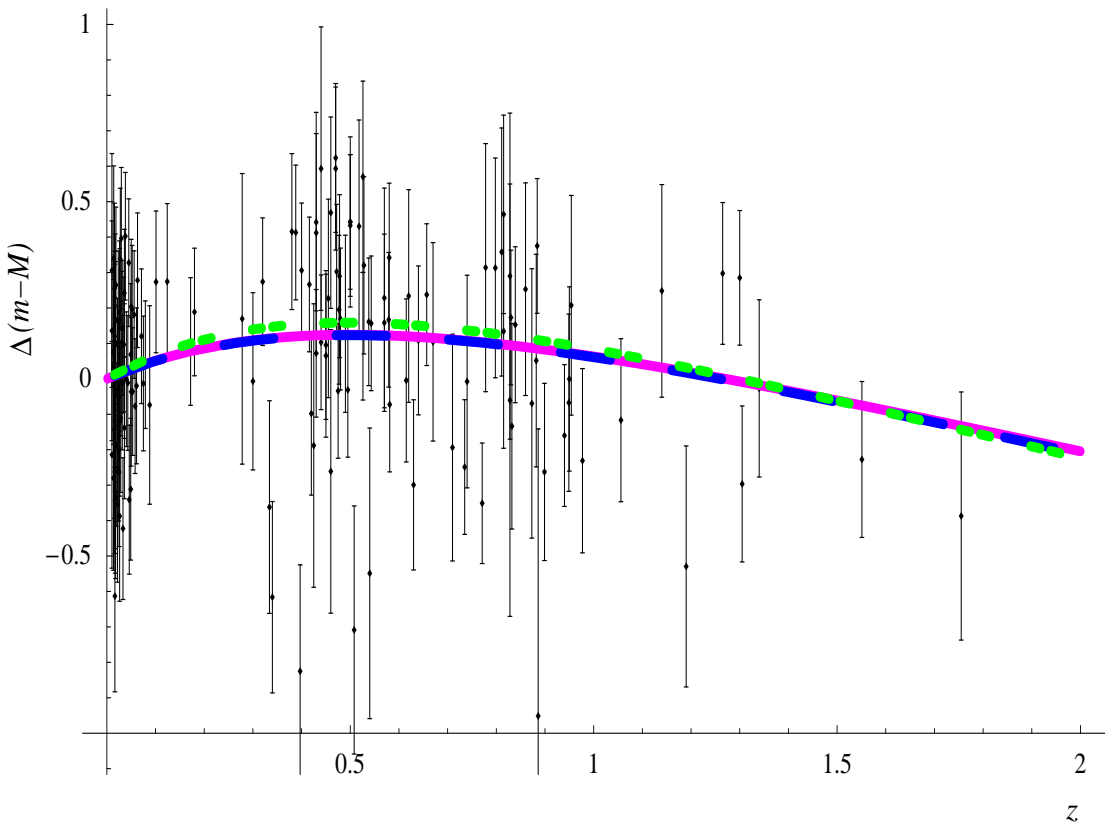}}
\caption{{\small The observed differential distance modulus of
high-redshift SNe from~\cite{Riess} relative to an empty universe
together with with the the curves for three models: a cosmological
constant with $\Omega_m=0.3$ (blue dashed curve), a cosmological
constant with $\Omega_m=0.35$ plus photon-axion oscillations with
$(L_{dec} H_0)^{-1} =0.25$ (purple solid curve) and a phantom
matter with $w=-1.25$ and  $\Omega_m=0.35$ (green double-dashed
curve). The three curves are practically all indistinguishable,
and fit the data equally well. \label{fig:bestfit}}}
\end{figure}

So let us consider the combined effect of dark energy with $w
\simeq -1$ and the photon-axion conversion. We can compare the
Hubble diagrams for SNe in the universe with and without
photon-axion mixing, and also with the universe dominated by dark
energy with a phantom-like equation of state. Assuming spatial
flatness and taking $\Omega_m=0.3$ and $\Omega_\Lambda=0.7$, in
Fig. (\ref{fig:bestfit}) we have plotted Hubble diagrams for the
case of cosmological constant without photon-axion conversion
(blue dashed line). Almost exactly underneath it we see the case
when $\Omega_m=0.35$, $\Omega_\Lambda=0.65$ and photon-axion
conversion with $(L_{dec} H_0)^{-1} =0.25$ (purple solid line).
Finally we plot $\Omega_m=0.35$ and a phantom with $w = -1.25$
(green double dashed line). We have also plotted the measured
differential distance moduli from the 157 Type IA SNe reported
recently in the ``gold sample'' of~\cite{Riess} (relative to an
empty universe corresponding to $\Delta(m-M)=0$), which include
the most distant SNe discovered using the Hubble Space Telescope.
We can see from Fig.~(\ref{fig:bestfit}) that the curves based on
these scenarios are almost indistinguishable from each other, and
thus each fit the observations equally well. Therefore, we
conclude that the combined effect of cosmological constant and
photon-axion conversion is practically indistinguishable from a
phantom with $w <-1$. To get an idea of what the value of the
best-fit ``phantomlike'' dark energy would be, we have compared
the differential distance modulus vs. redshift curves of the
cosmological constant plus axion system with those of some
phantomlike dark energy, and minimized the distance between these
curves. In this way we obtain an approximate expression for the
value of $w$ that a cosmological constant plus photon-axion
oscillations would fake:
\begin{equation}
w^{fake}\simeq -1 -(2.13\, \Omega_m +0.04) (L_{dec} \,H_0)^{-1}
~~.
\end{equation}

In order to find out which of these fake values of $w$ can occur
we have performed a fit to the gold sample of the SNe luminosity
\begin{figure}[thb]
\centerline{\includegraphics[width=17cm]{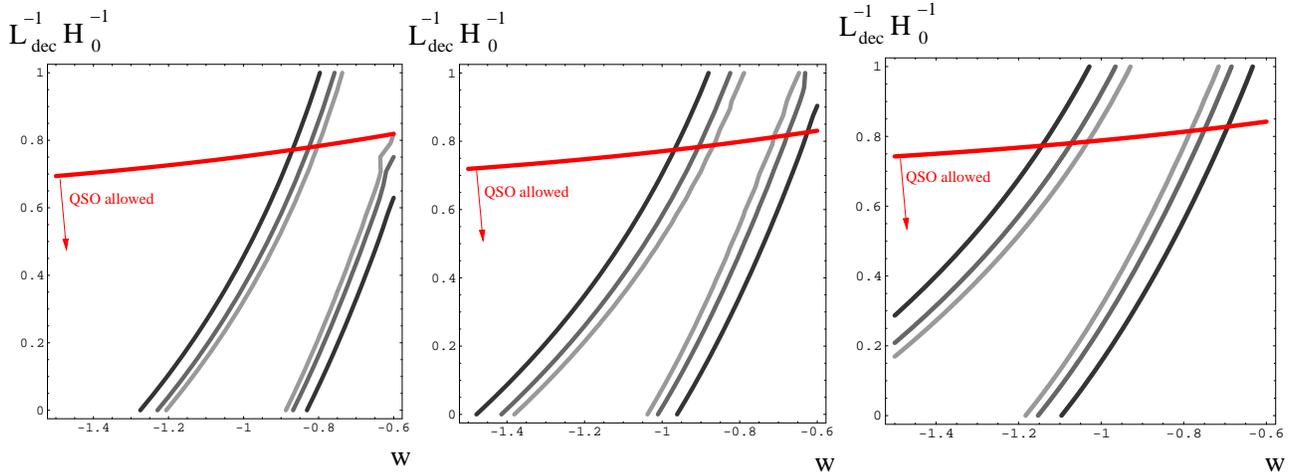}}
\begin{center}
\caption{\small The regions allowed by the latest SN
observations~\cite{Riess} at 90, 95 and 99 \% confidence level in
the $w-L_{dec}$ plane. The red curve shows the region allowed by
the study of the energy dependence of quasar spectra~\cite{mogoo}.
The left panel corresponds to $\Omega_m=0.3$, the middle one to
$\Omega_m=0.35$ while the right one to $\Omega_m=0.4$.
\label{fig:SNfit}}
\end{center}
\end{figure}
distances in~\cite{Riess}. In the first fit we have not restricted
the dark energy component to be a cosmological constant, but
rather allowed for a generic constant equation of state $w$. We
then varied the photon-axion oscillation decay length $L_{dec}$
and plotted the contours of constant $\chi^2$'s in
Fig.~(\ref{fig:SNfit}) for several values of $\Omega_m$. Note,
that for lower values of $\Omega_m$, the case of a cosmological
constant without photon-axion conversions is close to providing the
best fit to the data, while for larger values of $\Omega_m$ this
point is not even within the allowed region. The minimal $\chi^2$
values for all cases are around 176 for the 157 data points
from~\cite{Riess}. We have also demarcated the bound on the decay
length coming from the study of the energy dependence of the
quasar spectra from~\cite{mogoo}.

\begin{figure}[thb]
\centerline{\includegraphics[width=7cm]{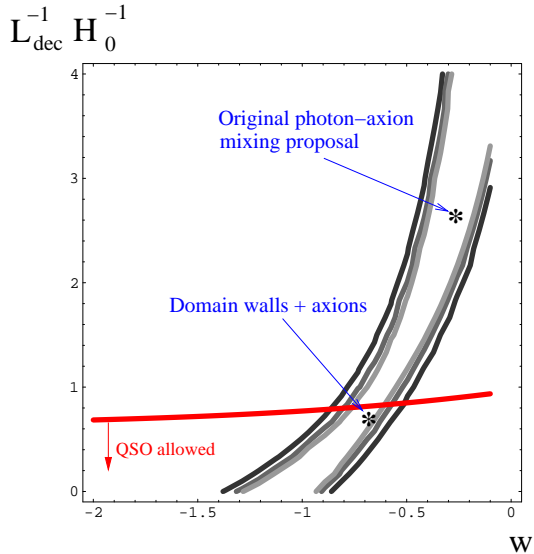}}
\begin{center}
\caption{\small The same as in Fig.~\ref{fig:SNfit} for $\Omega_m
=0.35$, except the region for a larger region of $w$ and
$L_{dec}$. This plot illustrates the fact that the original
photon-axion mixing proposal~\cite{ckt} would still fit all the
SNe observations (and also the WMAP constraints). The only real
constraints on the model that disfavor it as an explanation for
the majority of the dimming come from the study of the energy
dependence of quasar spectra~\cite{mogoo} or from violations of
reciprocity~\cite{baku}. The red curve shows the approximate bound
from the quasar studies. \label{fig:fullSNfit}}
\end{center}
\end{figure}
In Fig.~(\ref{fig:fullSNfit}) we have repeated the same analysis
except for a bigger region of the $w-L_{dec}$ parameter space, for
$\Omega_m=0.35$. The point of this analysis is to demonstrate that
the originally proposed mechanism for explaining all the SNe
dimming ($w=-1/3, L_{dec}\sim 0.5 H_0^{-1}$) still fits
the SNe observations very well. The only arguments disfavoring the case
of
$w=-1/3$ combined with photon-axion conversion as the source of
all the dimming are the bounds from~\cite{mogoo,baku}. If for any
reason these bounds would turn out to be unreliable (which is
likely the case with~\cite{baku}, given~\cite{uam}) then even the
original proposal would still be a viable explanation. More
importantly, we can see both from this figure and from
Fig.~(\ref{fig:SNfit}) that the case of a domain wall~\cite{Maxim}
equation of
state ($w=-2/3$) combined with photon-axion conversion is still
well within the allowed region for lower ($0.3-0.35$) values of
$\Omega_m$.

\begin{figure}[thb]
\centerline{\includegraphics[width=7cm]{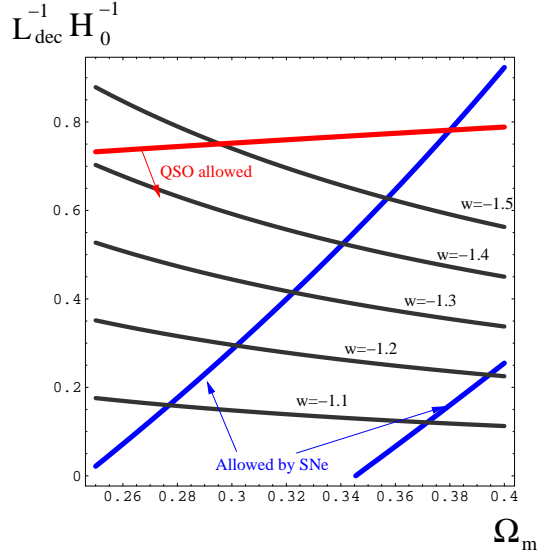}}
\begin{center}
\caption{\small The region allowed by the latest SN
observations~\cite{Riess} at the 95 \% confidence level in the
$\Omega_m -L_{dec}$ plane (blue curves). The equation of state of
the dark energy is fixed to be $w=-1$. The red curve again shows
the bound from the quasar spectra, while the black contours show
the approximate effective values of the equation of state one
would obtain by trying to fit the luminosity-distance curves. We
can see that values as low as $-1.5$ can be obtained for
$w^{fake}$. \label{fig:wommat}}
\end{center}
\end{figure}
Finally, we have scanned the parameter space in $\Omega_m$ and
$L_{dec}$, fixing the equation of state of the dark energy to be
that of a cosmological constant, $w=-1$. The allowed region
together with the approximate ``fake values'' of $w$, which one
would attribute to this system, are shown in
Fig.~(\ref{fig:wommat}). We can see from
this plot that with this simple system effective values of $w$ as
low as $-1.5$ can be obtained, without violating any experimental
bounds.

To conclude, we have considered the effect of conversion of
photons into axions, which evade detection on earth, in a universe
dominated by a cosmological constant on the dimming of distant
supernovae. The supernovae appear dimmer than they would with a
cosmological constant alone, and the effect turns around at
redshifts $z > 1$ just like any dark energy, thanks to the
saturation of the photon conversion into axions at  a third of the
initial flux. Therefore the combined effect of conversion of
photons into axions and a cosmological constant may appear like a
phantom dark energy with $w < -1$ to an observer bent on
interpreting the observations of dimming of SNe by dark energy
alone. The combination of photon-axion conversion with a
cosmological constant also makes it much easier to get within the
axion observational bounds, since it reduces the axion
contribution relative to the case which we have studied
previously, with dark energy obeying $w=-1/3$ \cite{ckt}. Our
model does not involve any ghost-like degrees or freedom, and the
Hamiltonian of the system remains bounded from below, guaranteeing
vacuum stability. Thus the usual effective field theory in curved
space applies and there are no problems with the instabilities
plaguing theories with phantoms and (uncondensed) ghosts. We
should stress that our mechanism only affects bright sources, and
electromagnetic radiation at frequencies above the microwave, and
so the observations involving CMB and large scale structure would
remain unaffected and fully consistent with cosmological constant
as dark energy. This is unlike the attempts to include $w<-1$
impersonators in the gravitational sector, which would affect
everything in the same way as long as one maintains the principle
of equivalence. Thus any disparity between different observations
of the equation of state parameter $w$ could be a very strong
signature of our mechanism. Additional signatures were discussed
in \cite{ckt,cktplasma,jecg,others,mbg,mogoo,baku}, and this host
of signals makes the mechanism amenable to verification and
constraints. The axions might even lead to super-GZK cosmic rays
generated by photon primaries \cite{gzk}. In light of this we
believe that an exploration of ultra-light axion effects in
cosmology is a profitable endeavor. Either such axions are really
there, and may be within observer's reach, or we may achieve the
strongest bounds on the ultra-light axions available today.

\section*{Acknowledgements}

We thank A. Albrecht, J. Frieman, G. Gabadadze, M. Kaplinghat, E.
Linder, A. Lue, R. Scoccimarro and L. Sorbo for useful
discussions. NK thanks the Aspen Center for Physics for kind
hospitality during the course of this work. The work of CC is
supported in part by the DOE OJI grant DE-FG02-01ER41206 and in
part by the NSF grants PHY-0139738  and PHY-0098631. The work of
NK is supported in part by the DOE Grant DE-FG03-91ER40674, in
part by the NSF Grant PHY-0332258 and in part by a Research
Innovation Award from the Research Corporation. J.T. is supported
by the U.S. Department of Energy under contract W-7405-ENG-36.


\end{document}